\documentclass{pasj00}

\begin{document}
\SetRunningHead{M. Uemura et al.}{Rapid optical fluctuations in V4641 Sgr}
\Received{2002/07/19}
\Accepted{2002/08/07}

\title{Rapid Optical Fluctuations in the Black Hole Binary, V4641 Sgr}

\author{
Makoto \textsc{Uemura},\altaffilmark{1}
Taichi \textsc{Kato},\altaffilmark{1}
Ryoko \textsc{Ishioka},\altaffilmark{1}
Kenji \textsc{Tanabe},\altaffilmark{2}
Seiichiro \textsc{Kiyota},\altaffilmark{3}\\
Berto \textsc{Monard},\altaffilmark{4}
Rod \textsc{Stubbings},\altaffilmark{5}
Peter \textsc{Nelson},\altaffilmark{6} 
Tom \textsc{Richards},\altaffilmark{7}
Charles \textsc{Bailyn},\altaffilmark{8}\\ and
Roland \textsc{Santallo},\altaffilmark{9}
}

\altaffiltext{1}{Department of Astronomy, Faculty of Science, Kyoto University,
  Sakyou-ku, Kyoto 606-8502}
\email{uemura@kusastro.kyoto-u.ac.jp}
\altaffiltext{2}{Department of Biosphere-Geosphere Systems, Faculty of 
Informatics, \\Okayama University of Science, Ridaicho 1-1, Okayama
700-0005}
\altaffiltext{3}{Variable Star Observers League in Japan (VSOLJ);
  Center for Balcony Astrophysics, \\1-401-810 Azuma, Tsukuba 305-0031}
\altaffiltext{4}{Bronberg Observatory, PO Box 11426, Tiegerpoort 0056,
  South Africa} 
\altaffiltext{5}{19 Greenland Drive, Drouin 3818, Victoria, Australia}
\altaffiltext{6}{RMB 2493, Ellinbank 3820, Australia}
\altaffiltext{7}{Woodridge Observatory, 8 Diosma Rd, Eltham, Vic 3095,
  Australia} 
\altaffiltext{8}{Department of Astronomy, Yale University, P.O. Box
  208101, New Haven, CT 06520-8101} 
\altaffiltext{9}{Faaa, Tahiti, French Polynesia}


%


\maketitle

\begin{abstract}
 We report on unprecedented short-term variations detected in the
 optical flux from the black hole binary system, V4641 Sgr.  Amplitudes
 of the optical fluctuations were larger at longer time scales, and
 surprisingly reached $\sim 60\%$ around a period of $\sim$10 min.  The
 power spectra of fluctuations are characterized by a power law
 ($\propto f^{-\alpha},\; \alpha \sim -1.7$).  It is the first case in
 black hole binaries that the optical emission was revealed to show
 short-term and large-amplitude variations given by such a power
 spectrum.  The optical emission from black hole binaries is generally
 dominated by the emission from the outer portion of an accretion disc.  
 The rapid optical fluctuations however indicate that the emission from
 an inner accretion region significantly contributes to the optical
 flux.  In this case, cyclo-synchrotron emission associated with various
 scales of magnetic flares is the most promising mechanism for the
 violently variable optical emission. 
\end{abstract}

\section{Introduction}
 
  In many steller-mass black holes, the X-ray flux rapidly oscillates in
various time-scales (\cite{lew95XB}; \cite{wei97CygX1};
\cite{rut99BHCQPO}).  The short time-scale variations are generally
believed to originate from the inner accretion flow onto a black hole.
Hence, X-ray fluctuations have received much attention for investigating
the physics of the accretion flow and the black hole itself
(\cite{che94BXCQPO}; \cite{wei98BHspin}; \cite{hom01J1550};
\cite{bai01nature}).   

  Here we report on short-term variations detected in the optical range, 
at which black hole binaries had been believed to be relatively calm. 
The object is called as V4641 Sgr ($V=13.8\;{\rm mag}$ 
at quiescence).  It forms a binary system with an orbital period of 2.8
d, containing a normal star of 5--8 solar-mass ($M_{\odot}$) and a black
hole of $\sim 10 M_{\odot}$ which accretes the overflowing gas from the
normal star (\cite{oro01v4641sgr}).  

This object first received much attention in 1999 September, when it
experienced a quite luminous, but unexpectedly short outburst
(\cite{uem02v4641}; \cite{smi99v4641}).  Another noteworthy feature of
this system is highly relativistic jets which were associated with this
outburst (\cite{hje00v4641}).  Radio observations detected superluminal
motion of jets with an apparent velocity of $\gtrsim 9.5c$
(\cite{hje00v4641}; \cite{oro01v4641sgr}).  These features make V4641
Sgr an outstanding black hole X-ray transient.  The mechanisms of its
outburst and highly relativistic jet are poorly understood.  

The next major outburst occurred in 2002 May, during which we detected
rapid optical fluctuations (\cite{mar02iauc}). 

\begin{figure*}
  \begin{center}
    \FigureFile(170mm,170mm){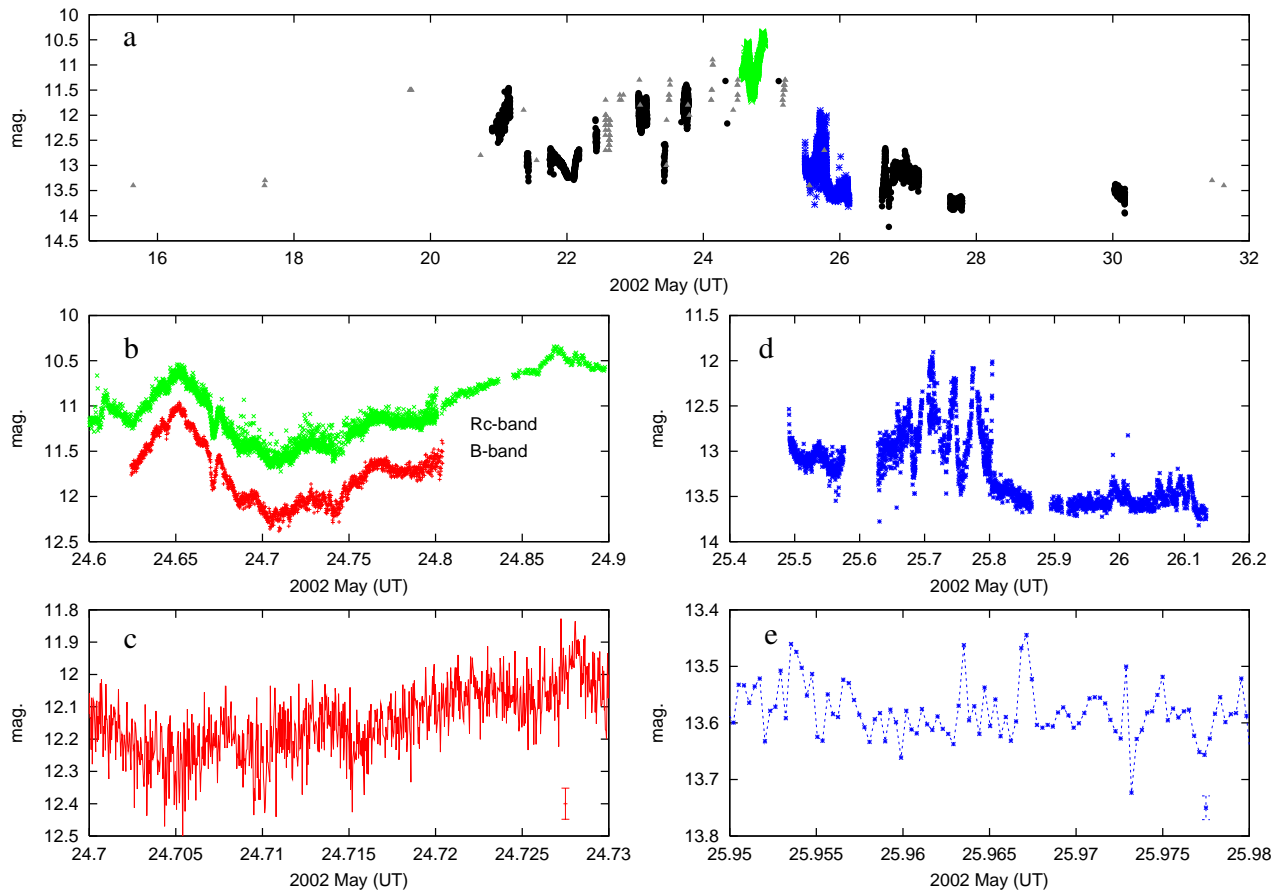}
  \end{center}
  \caption{Light curves of V4641 Sgr during the outburst in 2002 May.  a:
The whole light curve of the outburst.  The abscissa and ordinate denote
the date and $R_{\rm c}$ magnitude, respectively.  The gray
triangles denote the visual estimation reported to Variable Star Network
(VSNET; http://www.kusastro.kyoto-u.ac.jp/vsnet/).  The black, green,
and blue points denote our CCD time-series observations.  b, c, d, and
e: Light curves on May 24 (b, c) and 25--26 (d, e) showing short-term 
fluctuations.  Errors of the points are 0.02--0.05 mag in these
 time-series observations, and their typical errors are indicated in
 Figure 1c and 1e.}
\label{fig:lc02}
\end{figure*}

\section{Observation}

Our CCD photometric time-series observations were performed typically
with 30-cm class telescopes at Kyoto, Ouda, Okayama, Tsukuba,
Tiegerpoort, Ellinbank, and Eltham.  We calculated the differential
magnitude with a neighbour comparison star using CCD images which were  
subtracted the dark current image and performed flat fielding.  The
exposure times were 1--20 s.  The unfiltered-CCD magnitude scales of
each observatory were adjusted to that of the $R_{\rm c}$ system
obtained from $R_{\rm c}$-band observations at Ouda.  We neglect
possible small differences of variations between unfiltered CCD and
$R_{\rm c}$ systems since the sensitivity peak of the CCD camera is near
that of the $R_{\rm c}$ system and the color of the object is $B-V\sim
0$.  Heliocentric corrections to the observed times were applied before
the following analysis. 

\section{Result}

The outburst was first detected at 11.5 mag on May 19.699 UT.  Figure 1
shows the optical light curves of V4641 Sgr during the outburst in 2002
May.  Figure 1a is the whole light curve during the outburst.   As can
be seen in this figure, the object experienced repetitive brightening
with amplitudes of $\sim 1\;{\rm mag}$ and short durations of $\lesssim
1\;{\rm d}$ until May 25.   

As mentioned above, we detected rapid optical fluctuations during this
outburst.  Figure 1b shows an example of them observed around the
optical peak of the 2002 outburst.  We can see fluctuations of
brightness with amplitudes of 0.5--1.5 mag and time-scales of hours in
this figure.  Our $R_{\rm c}$ and $B$-band simultaneous observations
revealed that the amplitude of a large hump around May 24.65 is larger
in $B$ magnitude than that in $R_{\rm c}$ one.  This means that the
object became bluer when it reached the peak of the hump.  Figure 1d
also shows violent fluctuations with amplitudes of about 1 mag, however
the object suddenly became calm since May 25.8.  Such a dramatic
transition from the burst phase to the calm phase is analogous to X-ray
variations observed in one of the most famous black hole binaries, the
microquasar GRS 1915+105 (\cite{rao00grs1915QPO};
\cite{gre01grs1915donor}).  These hours- and minutes-order fluctuations 
appeared in all light curves taken with our CCD time-series observations
during the outburst (May 21--27).  The profiles of hours-order
variations were apparently variable with time, as can be seen in Figure
1b and 1d.   

The object was active even in shorter time-scales.  In Figure 1c, we
show $B$-band flux variations of an order of seconds--minutes.  Similar
fluctuations were observed even after the object faded from the
outburst, as shown in Figure 1e.  Compared with the hour-order
variations, they have smaller amplitudes of 0.1--0.2 mag, however the
errorbars indicated in Figure 1c and 1e are small enough to demonstrate
that they are real variations.  This observation makes the first
unambiguous detection of these time-scale fluctuations in the optical
flux from a black hole binary system (\cite{kan01nature};
\cite{cam97GX339}). 

Figure 2 shows power spectra of these fluctuations.  The power spectra
in this figure were calculated using the $B$-band on May 24 (as shown in
Figure 1b and 1c) and 25.  As can be seen in Figure 2, the photon noise
dominates at frequencies higher than 0.01 Hz, however the object has 
significant power in lower frequencies.  This demonstrates that
short-term fluctuations shown in Figure 1 are real variations at least,
in frequencies lower than 0.01 Hz, or periods longer than 100 s.   
The two power spectra have almost same features, while their difference
is somewhat larger at the lowest frequency region.  It is probably
caused by daily variations of hours-scale modulations.  The difference
in the low-frequency region suggests that different mechanisms worked
between hour-order, $\sim 1$ mag variations (Figure 1b and 1d) and
second--minute order, $< 1$ mag fluctuations (Figure 1c and 1e).    
The low-frequency part (0.0006--0.006 Hz) in the power spectra
correspond to the latter, shorter-term fluctuations, which was described
by a power law ($\propto f^{-\alpha}$).  Our $\chi^2$ fitting yielded
$\alpha=1.69\pm 0.20$ both in the case on May 24 and 25.    

\begin{figure}
  \begin{center}
    \FigureFile(80mm,80mm){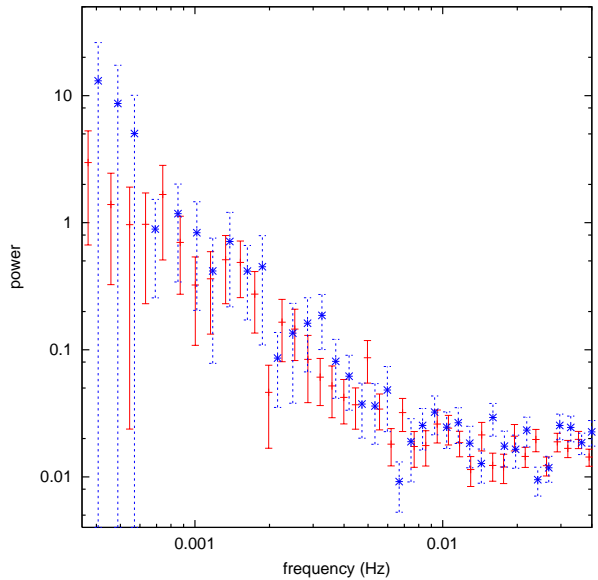}
  \end{center}
  \caption{Power spectrum of fluctuations observed on May 24 and 25.  The
abscissa and ordinate denote the frequency in Hz and the power in an
arbitrary unit.}
\label{fig:powspec}
\end{figure}

\section{Discussion}

It is quite unusual that short-term and large-amplitude variations, as 
shown in Figure 1, were detected in the optical range which is a 
long wavelength region compared with the X-ray range.  In general, the
optical emission from black hole binaries is considered to be the thermal
emission from the outer, low temperature portion of the disc
(\cite{hyn98J1655multiwavelength}; \cite{hyn02j1859}).  To date,
ordinary black hole binary binaries sometimes showed the optical-flux
variations, called ``superhumps'' (\cite{odo96BHXNSH}).  They can
however be explained by modulations at the tidally-distorted outer
accretion disc (\cite{has01BHXNSH}; \cite{whi88tidal}).  We can
naturally understand that variations originated from relatively low
temperature region are detected in the optical range.   

On the other hand, the optical emission in V4641 Sgr during the 2002
outburst has completely different features which were never observed in such
ordinary systems.  The time-scale of fluctuations is much shorter than
the orbital period of V4641 Sgr ($\sim 2.8$ d) and at the outermost part
of the accretion disc.  They hence indicate that the optical emission
originates from the inner region of the accretion disc where the matter
orbits with shorter time-scales and in high-temperature.  The striking
similarity to X-ray modulations in GRS 1915+105 also reminds us of
inner, short-time scale events.   

Another noteworthy feature during the 2002 outburst of V4641 Sgr is an
extremely small X-ray/optical flux ratio.  Observations with Rossi X-ray
Timing Explorer (RXTE) revealed that the object was active also in X-ray
during the optical outburst and reached a peak of about 60 mCrab on May
24.55 \footnote{$\langle$http://lheawww.gsfc.nasa.gov/users/swank/v4641sgr/$\rangle$}
(\cite{mar02iauc}).  The X-ray/optical flux ratio of V4641 Sgr is
calculated to be an order of 1, although those of typical systems are
$\sim 500$ (\cite{tan96XNreview}).  Such a small X-ray/optical flux
ratio means a peculiar energy spectrum in the X-ray--optical range, in
which the optical emission is exceptionally high.   

The extremely small X-ray/optical flux ratio indicates weak contribution
of thermal emission from an optically-thick disc.  Instead of thermal
disc emission, the synchrotron emission might be acceptable for a
dominant source of the optical emission.  During the outburst, the
inverted spectrum was reported by radio observations, which strongly 
indicates a strong contribution from synchrotron emission in the radio
range \footnote{$\langle$http://www.atnf.csiro.au/people/rsault/astro/v4641/$\rangle$}
\footnote{$\langle$http://vsnet.kusastro.kyoto-u.ac.jp/vsnet/Mail/vsnet-campaign-v4641sgr/msg00051.html$\rangle$}.
It is possible that that the strong synchrotron emission dominates even
in the infrared--optical range.  Under magnetic fields $B\sim
10^6$--$10^7$ G, theoretical calculations indicate that the synchrotron
emission from accretion discs can be its maximum at the optical range,
and furthermore, simulated energy spectra can yield extremely small
X-ray/optical ratio (1--10), as in V4641 Sgr (\cite{mer00j1118};
\cite{mar01j1118}).

The observed colours also support a significant contribution of
synchrotron emission at the optical range.  Our multi-colour photometry
on May 24 yielded de-reddened colours of the highly variable component of  
$B-V=-0.177\pm 0.039$, $V-R_c=-0.092\pm 0.039$, and $R_c-I_c=0.004\pm
0.026$ ($E(B-V)=0.32$ assumed; \cite{oro01v4641sgr}).  It should be
noticed that the colour of $R-I$ is exceptionally red compared with the
$B-V$ and the $V-R_c$.  This means that the $I_c$-band flux is
considerably in excess of that expected from the $B-V$ and the $V-R_c$.
Such a energy spectrum cannot be explained only with thermal disc
emission.  The synchrotron emission can be a strong candidate to
reconcile these colours.   

The above scenario can moreover provide a consistent picture for
short-term fluctuations.  The paradigm of cyclo-synchrotron emission
from magnetic flares has been recently discussed for a strong candidate
of short-term fluctuations in the black hole binaries (\cite{mer00j1118};
\cite{kan01nature}).  Various scales of magnetic flares in accretion discs
can produce various scales of flux variations of the cyclo-synchrotron
emission.  The fluctuations from such flares are reported to produce
power spectra with a power law of $\alpha\sim 2$ (\cite{min95lowstatedisk};
\cite{kaw00BHADfluctuation}).  This is acceptable for $\alpha$ observed
in X-ray emission from ordinary black hole binaries ($\alpha=1$--$2$)
(\cite{lew95XB}; \cite{vanderkli89QPOreview}), and also, in optical
emission from V4641 Sgr.  The cyclo-synchrotron emission mentioned above
is thus the most promising interpretation of the optical
seconds--minutes order fluctuations in V4641 Sgr.  While radio
observations showed no evidence of short-term fluctuations
\footnote{$\langle$http://vsnet.kusastro.kyoto-u.ac.jp/vsnet/Mail/alert7000/msg00351.html$\rangle$},
it can be naturally understand by considering emission sources different
from the optical one, for example, ejected clouds or optically-thick 
post-shock jet (\cite{mar01j1118}).

The synchrotron emission from jets is also a possible source of the
rapid optical variations although no evidence of jets has been reported
during the outburst in 2002 May (\cite{mir98grs1915}).  V4641 Sgr is
proposed to be a ``microblazar'' rather than a ``microquasar'' since the
jets observed during the 1999 outburst have been proposed to have a
largest bulk Lorentz factor of $Gamma\gtrsim 9.5$ among known galactic
sources (\cite{oro01v4641sgr}; \cite{mir99jet}).  If the large $Gamma$
is caused by a low inclination of the system, the rapid, large-amplitude
fluctuations in the 2002 May outburst may be explained with the Doppler
boosting in the jets.  In this case, however, the lack of radio
short-term fluctuations may be a more serious problem since we may need
another synchrotron source which should be steady and dominant at the
radio range.  

A noteworthy advantage of V4641 Sgr is its apparent brightness
($V=13.8$) which we can easily observe even with small telescopes,
compared with the other objects, for example, GRS 1915+105 ($V>19$) and
XTE J1118+480 ($V=18$--$19$).  In a number of black hole binaries, the
optical--UV flux is heavily obscured by intersteller extinction.  On the
other hand, it can be the most essential wavelength when the synchrotron
emission is dominant because its peak frequency can lie at the optical
range (\cite{mer00j1118}; \cite{mar01j1118}).  V4641 Sgr is a unique 
object which shows fluctuations caused by the magnetic activity detected
in the optical range.  In future, simultaneous observations from X-ray
to radio will been performed and enable us to study the short-term
fluctuations in multi-wavelengths.  Our discovery of these new features
of V4641 Sgr will thus lead a revolutionary advance in our understanding
of the magnetic activity in black hole accretion discs. 

\vskip 3mm

We are pleased to acknowledge comments by T. Kawaguchi.  
We are grateful to many amateur observers for supplying their vital
visual CCD estimates via VSNET.  This work is partly supported by a
grant-in aid (13640239) from the Japanese Ministry of Education,
Culture, Sports, Science and Technology.  Part of this work is supported
by a Research Fellowship of the Japan Society for the Promotion of
Science for Young Scientists (MU).


\begin{thebibliography}{}

\bibitem[Bailyn(2001)]{bai01nature}
  Bailyn, C.\ 2001, \nat, 414, 499

\bibitem[Chen, Taam(1994)]{che94BXCQPO}
  Chen, X. \& Taam, R.~E.\ 1994, \apj, 431, 732

\bibitem[Greiner et~al.(2001)]{gre01grs1915donor}
  Greiner, J., Cuby, J.~G., McCaughrean, M.~J., Castro-Tirado, A.~J., \&
  Mennickent, R.~E.\ 2001, \aap, 373, L37

\bibitem[Haswell et~al.(2001)]{has01BHXNSH}
  Haswell, C.~A., King, A.~R., Murray, J.~R., \& Charles, P.~A.\ 2001, \mnras,
  321, 475

\bibitem[Hjellming et~al.(2000)]{hje00v4641}
  Hjellming, R.~M., Rupen, M.~P., Hunstead, R.~W., Campbell-Wilson, D.,
  Mioduszewski, A.~J., Gaensler, B.~M., Smith, D.~A., Sault, R.~J., {et~al.}\
  2000, \apj, 544, 977

\bibitem[Homan et~al.(2001)]{hom01J1550}
  Homan, J., Wijnands, R., van~der Klis, M., Belloni, T., van Paradijs, J.,
  Klein-Wolt, M., Fender, R., \& Mndez, M.\ 2001, \apjs, 132, 377

\bibitem[Hynes et~al.(2002)]{hyn02j1859}
  Hynes, R.~I., Haswell, C.~A., Chaty, S., Shrader, C.~R., , \& Cui, W.\ 2002,
  \mnras, 331, 169

\bibitem[Hynes et~al.(1998)]{hyn98J1655multiwavelength}
  Hynes, R.~I., Haswell, C.~A., Shrader, C.~R., Chen, W., Horne, K., Harlaftis,
  E.~T., O'Brien, K., Hellier, C., \& Fender, R.~P.\ 1998, \mnras, 300, 64

\bibitem[Kanbach et~al.(2001)]{kan01nature}
  Kanbach, G., Straubmeier, C., Spruit, H.~C., \& Belloni, T.\ 2001, \nat, 414,
  180

\bibitem[Kawaguchi et~al.(2000)]{kaw00BHADfluctuation}
  Kawaguchi, T., Mineshige, S., Machida, M., Matsumoto, R., \& Shibata, K.\
  2000, \pasj, 52, L1

\bibitem[Lewin et~al.(1995)]{lew95XB}
  Lewin, W. H.~G., van Paradijs, J., \& van~den Heuvel, E. P.~J.\ 1995, in
  X-ray Binaries, ed. Cambridge (Cambridge Univ. Press)

\bibitem[Markoff et~al.(2001)]{mar01j1118}
  Markoff, S., Falcke, H., \& Fender, R.\ 2001, \aap, 372, L25

\bibitem[Markwardt, Swank(2002)]{mar02iauc}
  Markwardt, C.~B. \& Swank, J.~H.\ 2002, \iaucirc, 7906

\bibitem[Merloni et~al.(2000)]{mer00j1118}
  Merloni, A., Di~Matteo, T., \& Fabian, A.~C.\ 2000, \mnras, 318, L15

\bibitem[Mineshige et~al.(1995)]{min95lowstatedisk}
  Mineshige, S., Kusnose, M., \& Matsumoto, R.\ 1995, \apjl, 445, L43

\bibitem[Mirabel et~al.(1998)]{mir98grs1915}
  Mirabel, I. F., Dhawan, V., Chaty, S., Rodriguez, L. F., Marti, J.,
		      Robinson, C. R., Swank, J., \& Geballe, T. R.\
		      1998, \aap, 330, L9 

\bibitem[Mirabel, Rodriguez (1999)]{mir99jet}
  Mirabel, I. F., \& Rodriguez, L. F.\ 1999, \araa, 37, 409

\bibitem[O'Donoghue, Charles(1996)]{odo96BHXNSH}
  O'Donoghue, D. \& Charles, P.~A.\ 1996, \mnras, 282, 191

\bibitem[Orosz et~al.(2001)]{oro01v4641sgr}
  Orosz, J.~A., Kuulkers, E., van~der Klis, M., McClintock, J.~E., Garcia,
  M.~R., Callanan, P.~J., Bailyn, C.~D., Jain, R.~K., \& Remillard, R.~A.\
  2001, \apj, 555, 489

\bibitem[Rao et~al.(2000)]{rao00grs1915QPO}
  Rao, A.~R., Naik, S., Vadawale, S.~V., \& Chakrabarti, S.~K.\ 2000, \aap,
  360, L25

\bibitem[Rutledge et~al.(1999)]{rut99BHCQPO}
  Rutledge, R.~E., Lewin, W. H.~G., van~der Klis, M., van Paradijs, J., Dotani,
  T., Vaughan, B., Belloni, T., Oosterbroek, T., \& Kouveliotou, C.\ 1999,
  \apjs, 124, 265

\bibitem[Smith et~al.(1999)]{smi99v4641}
  Smith, D.~A., Levine, A.~M., \& Morgan, E.~H.\ 1999, \iaucirc, 7253

\bibitem[Steiman-Cameron et~al.(1997)]{cam97GX339}
  Steiman-Cameron, T.~Y., Scargle, J.~D., Imamura, J.~N., \& Middleditch, J.\
  1997, \apj, 487, 396

\bibitem[Tanaka, Shibazaki(1996)]{tan96XNreview}
  Tanaka, Y. \& Shibazaki, N.\ 1996, \araa, 34, 607

\bibitem[Uemura et~al.(2002)]{uem02v4641}
  Uemura, M., Kato, T., Watanabe, T., Stubbings, R., Monard, B., \& Kawai, N.\
  2002, \pasj, 54, 95

\bibitem[van~der Klis(1989)]{vanderkli89QPOreview}
  van~der Klis, M.\ 1989, \araa, 27, 517

\bibitem[Wei et~al.(1998)]{wei98BHspin}
  Wei, C., Zhang, S.~N., \& Chen, W.\ 1998, \apj, 492, L53

\bibitem[Wei et~al.(1997)]{wei97CygX1}
  Wei, C., Zhang, S.~N., Focke, W., \& Swank, J.~H.\ 1997, \apj, 484, 383

\bibitem[Whitehurst(1988)]{whi88tidal}
  Whitehurst, R.\ 1988, \mnras, 232, 35

\end{thebibliography}

\end{document}